# Partial Discharge Direction-of-Arrival Estimation in Air-Insulated Substation by UHF Wireless Array and RSSI Maximum Likelihood Estimator

Bei Han [1], Lingen Luo [1,*], Gehao Sheng [1], Xiuchen Jiang [1]

[1] Department of Electrical Engineering, Shanghai Jiao Tong University, No. 800 Dongchuan Road, Shanghai, People's Republic of China
[*] llg523@sjtu.edu.cn

**Abstract:** The quick detection and localization of partial discharge (PD) in an air-insulated substation (AIS) based on ultrahigh-frequency (UHF) sensor arrays are efficient for power equipment monitoring. The adopted UHF PD time difference of arrival (TDOA) methods mainly use the time difference of electromagnetic wave signals. Thus, the system requires both a high sampling rate and time synchronization accuracy, leading to a high cost and large size. In this study, the feasibility and accuracy of PD DOA in an AIS were investigated using a UHF wireless sensor array and the received signal strength indicator. First, the power pattern of the designed UHF wireless sensor array was obtained via an offline experiment. Then, a statistical approach to the PD DOA method based on the maximum likelihood estimator was employed to obtain the preliminary DOA result. Finally, interpolation and clustering algorithms were used to improve the accuracy of the DOA. A laboratory test was conducted. The average error of the PD DOA was less than 6°, and the cost-effectiveness and portability were clearly improved.

## 1. Introduction

Partial discharge (PD) is one of the main reasons for dielectric aging and the decline of insulating properties for power equipment [1–3]. The detection and localization of PD using ultrahigh-frequency (UHF) array technology for the early warning of power equipment has been extensively studied [4]. UHF technology has the advantages of high sensitivity and fast propagation speed [5]; however, limited by economic considerations, online UHF PD monitoring is only used for such important devices as gas-insulated systems [6] and transformers [7]. Thus, a rapid PD detection and localization method for air-insulated substations (AISs) is proposed based on the UHF sensor array as a cost-effective solution [8–10]. One notable limitation of this system is that the PD localization algorithm used is mainly based on the time difference of the electromagnetic wave signal, i.e., the time difference of arrival (TDOA), which requires a high sampling rate (GSa/s) and high time synchronization accuracy (ns) [11–13]. Correspondingly, the high cost and large size of commercial systems limit its field application.

Recently, PD localization in air-insulated substations based on the UHF received signal strength indicator (RSSI) has been proposed and studied owing to its low hardware cost and better environmental adaptability [14]. The PD localization algorithms based on UHF RSSI are divided into two main categories. One is fingerprint based, which usually collects the fingerprint map by a site survey before the PD localization [14–16]. The PD source is localized by pattern recognition of the prebuilt RSSI fingerprint map. However, the requirement of the RSSI fingerprint map in the offline stage makes it difficult to deploy in substations [15]. In addition, when the environment changes, the fingerprint map needs to be rebuilt. The other is range-based PD localization, and several sensors are needed to collect the RSSI values. The PD UHF signal transmission distance to each sensor is calculated by the signal attenuation model, and the coordinate of the PD can be determined [17–19]. This method is significantly influenced by shadowing and multipathing effects, which make it infeasible for field applications in substations.

Therefore, in this study, the feasibility of PD DOA in AISs was investigated using a UHF wireless sensor array and RSSI values. A statistical approach based on the RSSI and maximum likelihood estimator (MLE) [20–22] is proposed. The basic idea is that the RSSI values are collected by the UHF sensor array, and the location of the PD source is estimated by the power pattern of the designed UHF sensor array and MLE. The likelihood function of MLE is a conditional probability density function with unknown parameters. The purpose of MLE is to use a known sample to deduce the parameters that maximize the likelihood function. Furthermore, interpolation and clustering algorithms are adopted for better PD DOA accuracy.

The remainder of this report is organized as follows. In Section 2, the designed UHF wireless sensor array with its power pattern is introduced. In Section 3, the specific algorithm of PD DOA based on RSSI values and MLE is described. The interpolation and clustering methods are also addressed. In Section 4, the laboratory tests are described, the corresponding results are given, and the error analysis of the proposed method is addressed. Section 5 provides a comparison of different PD localization methods. Section 6 concludes the report and highlights future work.

## 2. Framework of PD DOA Based on UHF Wireless Sensor Array

### 2.1. Power Pattern of Designed UHF Wireless Sensor Array

With the development of the Internet of things (IoT), there has been adoption of wireless technology for PD detection and monitoring devices to make systems lighter and easier to use. The UHF wireless sensor designed in this study is shown in Fig. 1. The sensor is composed of a printed circuit board UHF antenna and a corresponding signal-processing circuit. For better elimination of interference, the antenna and circuit are placed in different layers of the metal shell, as shown in Fig. 1(a).

Specifically, the input signal bandwidth is from 300 to 1500 MHz, which is the typical UHF frequency for PD measurement. The sample rate is as low as 2.3 MHz, and the wireless communication is according to the Wi-Fi protocol. The sensor is powered by a lithium battery, which makes it easy to use and convenient to carry. Typical measured RSSI data are shown in Fig. 1(b).

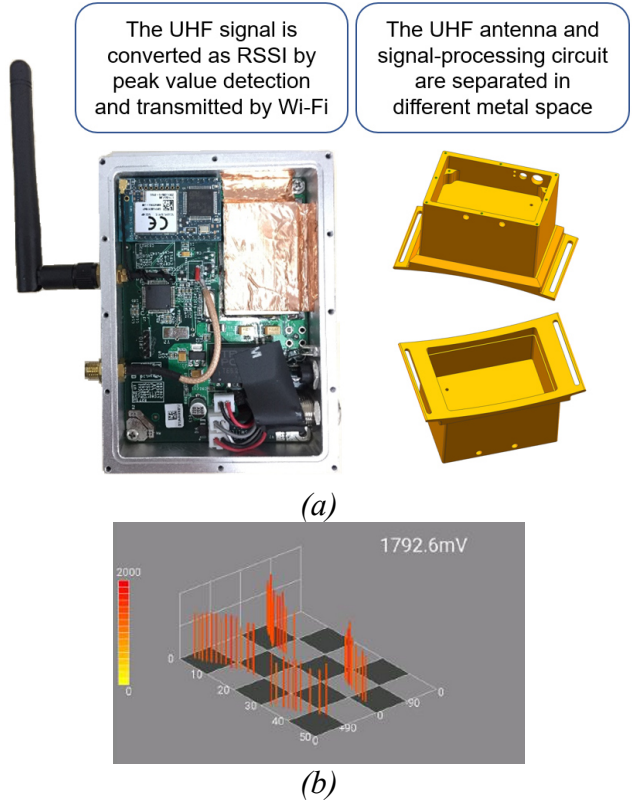

*Fig. 1. Designed UHF wireless sensor: **(a)** structure and assembly method and **(b)** typical measured RSSI values*

The proposed PD DOA method in AISs is performed by a UHF wireless sensor array, as shown in Fig. 2. The array is composed of four UHF wireless sensors and their antenna sides toward the outside. The four sensors are placed evenly, which means that the azimuth difference of each sensor is 90°. The RSSI values (in the form of voltages) measured by the sensor array are transmitted to the computer.

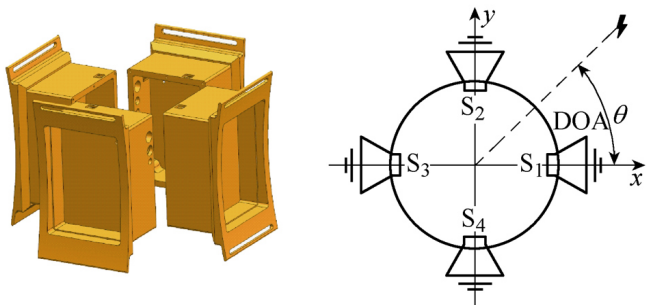

*Fig. 2. Diagram of UHF wireless sensor array*

For the PD DOA, the power pattern of the designed sensor array needs to be constructed. If the sensor array is taken as the centre of a circle, 18 points on a circle with a radius of 3 m are chosen as the test points. The incident azimuth (denoted as $\theta$, as shown in Fig. 2) of the 18 points are recorded as 10°, 30°, …, 350°.

An electrostatic discharge (ESD) gun that obeys the electromagnetic compatibility standard EN/IEC 61000-4-2 was used to generate a discharge pulse along the selected test points [23]. At each point, the ESD gun was used to generate a PD pulse 50 times, and the UHF sensors collected the corresponding RSSI values. When the RSSI values obtained by each sensor at the same point are averaged, the received power matrix of the sensor array can be denoted as

$$\boldsymbol{\Psi} = \begin{bmatrix} r_{1,1} & r_{1,2} & \cdots & r_{1,4} \\ r_{2,1} & r_{2,2} & \cdots & r_{2,4} \\ \vdots & \vdots & & \vdots \\ r_{18,1} & r_{18,2} & \cdots & r_{18,4} \end{bmatrix} \quad (1)$$

where each row of $\boldsymbol{\Psi}$ denotes the RSSI average values of the four UHF sensors at each point.

The matrix $\boldsymbol{\Psi}$ is normalised by

$$r_{i,j} = \frac{r_{i,j}}{\max r_{i,j}} \quad i = 1, 2, \cdots, 18 \quad j = 1, 2, 3, 4 \quad (2)$$

Then, the power pattern of the designed UHF wireless sensor array is obtained, as shown in Fig. 3.

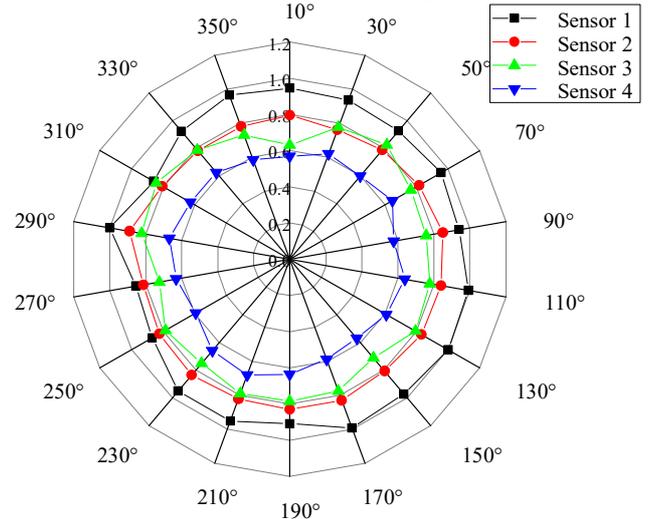

*Fig. 3. Power pattern of designed UHF sensor array*

### 2.2. Framework of Proposed PD DOA

A flow chart of the proposed PD DOA method for AISs based on RSSI MLE is shown in Fig. 4. The main processes are depicted as follows. First, the power pattern of the UHF sensor array used in the offline experiment is constructed. Then, the sensor array receives RSSI values when PD occurs. After several sampling times, the data processing, including normalization, MLE, and interpolation, is performed, and the preliminary DOA result is obtained. Furthermore, to improve accuracy, a clustering algorithm is used for multiple DOA results and gives the final DOA result.

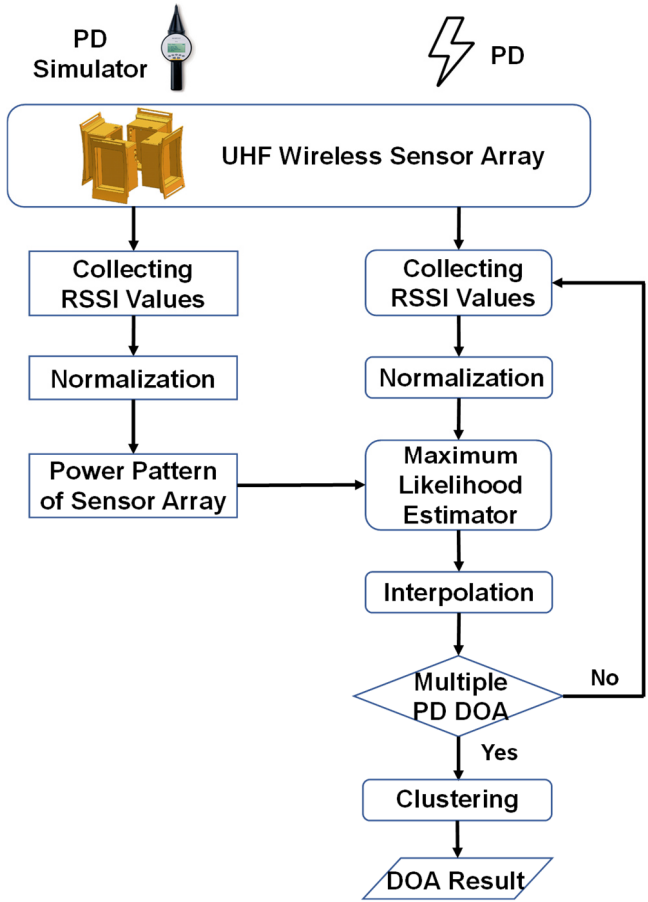

**Fig. 4.** *Flow chart of partial discharge direction of arrival estimation*

## 3. PD DOA Based on UHF RSSI MLE

### 3.1. Mathematics of UHF RSSI MLE

First, the relationship between the measured RSSI values and the average value of power computed from the received signal (RSSI values) are investigated. It is assumed that $M$ UHF sensors in the sensor array collect $K$ samples of the PD signal. The $k^{th}$ sampled signal $r_m(k)$ collected by the $m^{th}$ sensor is written as

$$r_m(k) = a_m(\theta)s(k) + n_m(k) \quad (3)$$

where $a_m(\theta)$ is the attenuation caused by the UHF sensor, $s(k)$ is the electromagnetic wave that arrives at the sensor, and $n_m(k)$ is white noise that follows the Gaussian distribution with mean value 0 and variance $\sigma^2$. Therefore, the average power of the $K$-samples PD signal of the $m^{th}$ sensor is

$$P_{r,m} = \frac{1}{K}\sum_{k=1}^{K}|r_m(k)|^2$$
$$= \frac{1}{K}\sum_{k=1}^{K}\left(a_m^2(\theta)s^2(k) + 2a_m(\theta)s(k)n_m(k) + n_m^2(k)\right) \quad (4)$$

Because the average of $n_m(k)$ is 0, the second item of the right part of Eq. (4) is also 0.

The transmitted PD signal power that arrives at the sensors can be written as

$$P_s = \frac{1}{K}\sum_{k=1}^{K}|s(k)|^2 \quad (5)$$

The power pattern of the $m^{th}$ sensor $g_m(\theta)$

$$g_m(\theta) = a_m^2(\theta) \quad (6)$$

The power patterns for all UHF sensors are denoted as $g(\theta)=[g_1(\theta), g_2(\theta),…, g_M(\theta)]$ and normalized by max $g(\theta)=1$. Thus, the signal-to-noise ratio (SNR) is defined as

$$\text{SNR} = \frac{P_s}{\sigma^2} \quad (7)$$

Substituting Eq. (5) and (6) into Eq. (4), one obtains

$$P_{r,m} = g_m(\theta)P_s + \frac{1}{K}\sum_{k=1}^{K}n_m^2(k) \quad (8)$$

The sum of the received PD power over $K$ samples is obtained by

$$S_{r,m} = KP_{r,m} = \sum_{k=1}^{K}|r_m(k)|^2 \quad (9)$$

where $S_{r,m}$ follows a noncentral $\chi^2$ distribution with $K$ degrees of freedom, denoted as $S_{r,m} \sim \chi^2(K, \lambda_m, \sigma^2)$. Because the RSSI values measured by each sensor are independent of each other, the average value of power computed from the received signal can follow the chi-square distribution [24]. The parameter $\lambda_m$ can be calculated by

$$\lambda_m = \sum_{k=1}^{K}E[r_m(k)]^2 = \sum_{k=1}^{K}a_m^2(\theta)|s(k)|^2$$
$$= Kg_m(\theta)P_s \quad (10)$$

Its probability density function (PDF) is calculated as

$$p_{S_{r,m}}(x) = \frac{1}{2\sigma^2}\left(\frac{x}{\lambda_m}\right)^{\frac{K-2}{4}} e^{-\frac{\lambda+x}{2\sigma^2}} I_{\frac{K}{2}-1}\left(\frac{\sqrt{\lambda_m x}}{\sigma^2}\right) \quad (11)$$

where $I_{\frac{K}{2}-1}(\frac{\sqrt{\lambda_m x}}{\sigma^2})$ is the modified Bessel function of the first kind.

The mean and variance of $P_{r,m}$ can be calculated using $S_{r,m}$:

$$\mu_m = E[P_{r,m}] = \frac{1}{K}E[S_{r,m}] = \frac{1}{K}(K\sigma^2 + \lambda_m)$$
$$= \sigma^2 + g_m(\theta)P_s \quad (12)$$

$$\sigma_m^2 = \text{VAR}[P_{r,m}] = \frac{1}{K^2}\text{VAR}[S_{r,m}]$$
$$= \frac{1}{K^2}(2K\sigma^4 + 4\sigma^2\lambda_m) = \frac{2}{K}(\sigma^4 + 2\sigma^2 g_m(\theta)P_s) \quad (13)$$

When $K$ or $\lambda_m$ is large, $P_{r,m}$ approximately follows a Gaussian distribution $P_{r,m} \sim N(\mu_m, \sigma_m^2)$. The corresponding PDF of $P_{r,m}$ is given by

$$p_{P_{r,m}} = \frac{1}{\sqrt{2\pi}\sigma_m}\exp\left(-\frac{(P_{r,m} - \mu_m)^2}{2\sigma_m^2}\right)$$
$$= \sigma^2 + g_m(\theta)P_s \quad (14)$$

Then, the maximum likelihood estimator [25] is adopted to infer the azimuth angle. For parameters of azimuth angle $\theta$, unknown signal power $P_s$, and noise variance $\sigma^2$, the estimation is given by

$$\Gamma = \left[\theta, P_s, \sigma^2\right] \quad (15)$$

The received signal power of $M$ sensors is recorded as $P_r=[P_{r,1}, P_{r,2},…, P_{r,M}]$; thus, the likelihood function is written as

$$p(P_r;\Gamma) = \prod_{m=1}^{M} p(P_{r,m};\Gamma) \quad (16)$$

The log likelihood function of Eq. (16) is

$$\ln p(P_r;\Gamma) = \sum_{m=1}^{M} \ln \frac{1}{\sqrt{2\pi}\sigma_m} \exp\left(-\frac{(P_{r,m}-\mu_m)^2}{2\sigma_m^2}\right)$$

$$= \sum_{m=1}^{M}\left(-\frac{1}{2}\ln(2\pi\sigma_m^2) - \frac{1}{2\sigma_m^2}(P_{r,m}-\mu_m)^2\right)$$

$$= \sum_{m=1}^{M} -\frac{1}{2}\ln\left(\frac{4\pi}{K}(\sigma^4 + 2\sigma^2 g_m(\theta)P_s)\right)$$

$$-\sum_{m=1}^{M} \frac{K(P_{r,m}-\sigma^2-g_m(\theta)P_s)^2}{4(\sigma^4+2\sigma^2 g_m(\theta)P_s)} \quad (17)$$

Therefore, the MLE can be calculated by

$$\hat{\Gamma}_{ML} = \arg\max_{\Gamma} \ln p(P_r;\Gamma)$$

$$= \arg\min_{\Gamma} \sum_{m=1}^{M} (\ln\left(\frac{4\pi}{K}(\sigma^4+2\sigma^2 g_m(\theta)P_s)\right) \quad (18)$$

$$+ \frac{K(P_{r,m}-\sigma^2-g_m(\theta)P_s)^2}{2\sigma^2(\sigma^2+2g_m(\theta)P_s)})$$

When Eq. (18) reaches its minimum, the corresponding $\theta$ is the DOA result of the PD.

### 3.2. Interpolation for Better Resolution

As the derivation in Section 3.1 shows, the accuracy of the PD DOA based on MLE depends on the resolution of the power pattern of the UHF sensor array. Therefore, for better resolution, spline interpolation was adopted in this study. The power pattern of UHF sensor array is measured with 20° intervals, namely (10°, 30°, …, 350°), as shown in the Fig. 3. Therefore, we use the interpolation method to increase the resolution from 20° interval to 1° interval.

Specifically, the 18 points (10°, 30°, …, 350°) to build the power pattern of the UHF sensor array were set as the abscissa values. The values of the likelihood functions calculated according to Eq. (18) are set as ordinate values. Then, spline interpolation in the range of 0°–360° is carried out as follows.

$$s(x) = a_j x^3 + b_j x^2 + c_j x + d_j \quad (19)$$

where $a_j$, $b_j$, $c_j$, and $d_j$ are unsolved coefficients, and $j$ denotes the $j$th interval of the abscissa.

When applying the interpolation method, first, the curve between two adjacent points is smoothed using Eq. (19). Then, the lowest point along the entire curve is found, and the angle value of the lowest point is taken as the preliminary DOA result. The difference between only MLE and MLE with interpolation is shown in Fig. 5. As a limit to the resolution of the power pattern, if only MLE is used, the minimum MLE value is the point in the red box. For MLE with interpolation, the minimum MLE value is the intersection of the blue line and black curve.

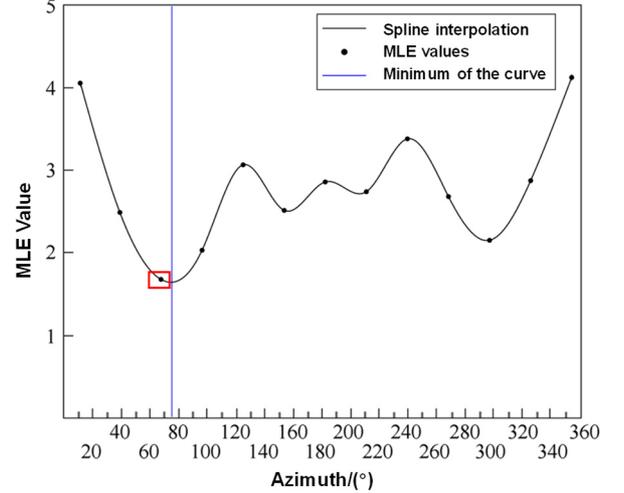

*Fig. 5. Comparison of only MLE and MLE with interpolation*

### 3.3. Clustering for Better Accuracy

Furthermore, considering that PD is a continuous process, the multiple PD DOA results calculated by MLE can be analysed by a clustering algorithm to improve the localization accuracy.

Considering that PD is a continuous process, clustering method is applied to analysis multiple PD DOA results for a better accuracy. K-means [26] as a typical clustering algorithm is adopted, since it is simple and suitable for the proposed method: it can obtain rather satisfying results without adding too much computation complexity. When the DOA results of $N$-sample PD have been calculated, K-means clustering divides the results into $k$ clusters by minimizing the Euclidean distance for all PD DOA results:

$$V = \sum_{i=1}^{k} \sum_{\theta_j \in M_i} (\theta_j - \mu_i)^2 \quad (20)$$

where $\mu_i$ is the initial cluster centre, and $V$ is the sum of the distance from all PD DOA results to their respective centre.

## 4. Laboratory Tests and Analyses

### 4.1. Test Setup and Results

To verify the feasibility and accuracy of the proposed method, a laboratory test was performed. The PD detection and orientation system based on the UHF wireless sensor array is shown in Fig. 6.

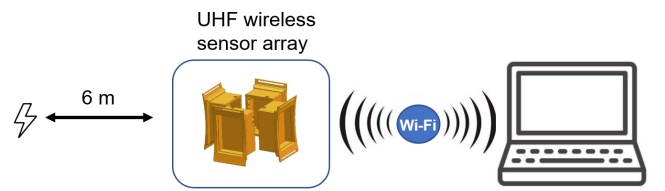

*Fig. 6. Diagram of UHF wireless sensor array PD detection and orientation system for AIS*

The sensor array was placed on a 1.2-m-high tripod. The PD simulator (Sparks Instrument, Switzerland) was used to generate the PD pulse. It was 6 m away from the sensor array. The experimental setup is shown in Fig. 7. When the PD pulses are generated, the excited electromagnetic wave signal radiates. The UHF signal was received and detected by the sensors, and then the RSSI values were transmitted to the computer through Wi-Fi. The PD DOA results were finally obtained by the proposed MLE method.

The azimuth angles of the PD were set as 40°, 80°, 150°, and 340°, and the PD pulse was generated 120 times at each azimuth angle for testing. Because of the characteristics of detection, some PD pulses could not be captured accurately by the UHF wireless sensor. Therefore, in each group of 120 tests, the true effective number of PD pulses was not the same.

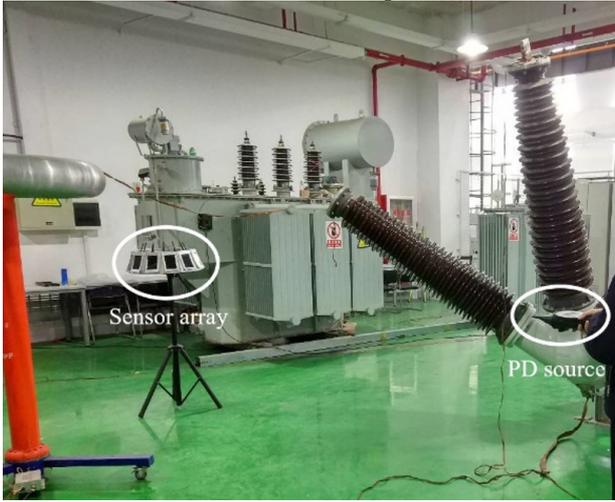

*Fig. 7.* Laboratory test environment

The PD DOA results for 80° obtained by MLE (the effective PD pulse number was 108) are shown in Fig. 8, while those obtained by both MLE and interpolation are shown in Fig. 9. The PD DOA results for 340° obtained by MLE (the effective PD pulse number was 117) are shown in Fig. 10. The results of both MLE and interpolation are shown in Fig. 11. Figs. 8 and 10 show that most PD DOA results were 90° (near 80°) and 330° (near 340°). This was because 90° and 330° are two selected points for the power pattern of UHF sensors. Therefore, the PD DOA results only by MLE will drop in the set of (10°, 30°, …, 350°). When interpolation was applied, the PD DOA resolution is increased to 1°, as shown in the subgraphs of Figs. 9 and 11.

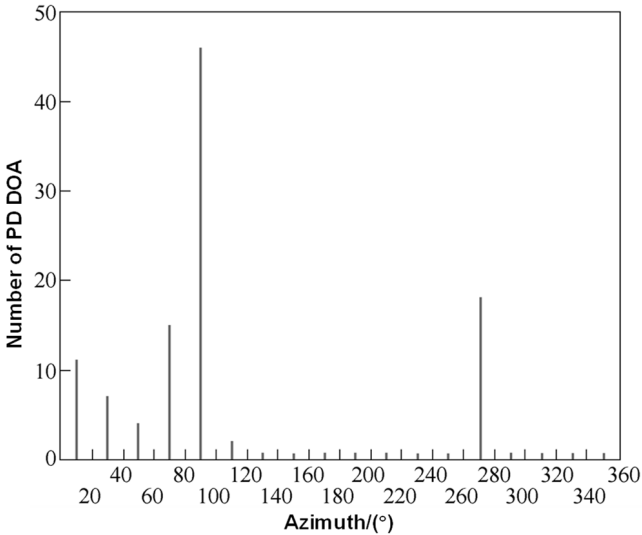

*Fig. 8.* Azimuth diagram of PD DOA results at 80° by MLE

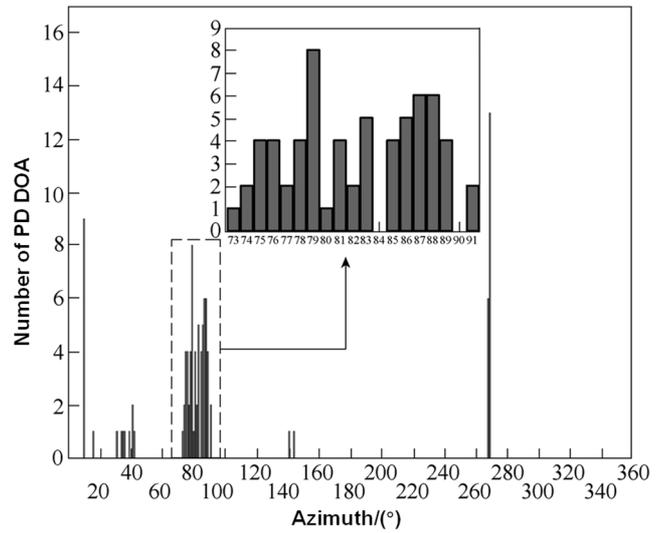

*Fig. 9.* Azimuth diagram of PD DOA results at 80° by MLE and spline interpolation

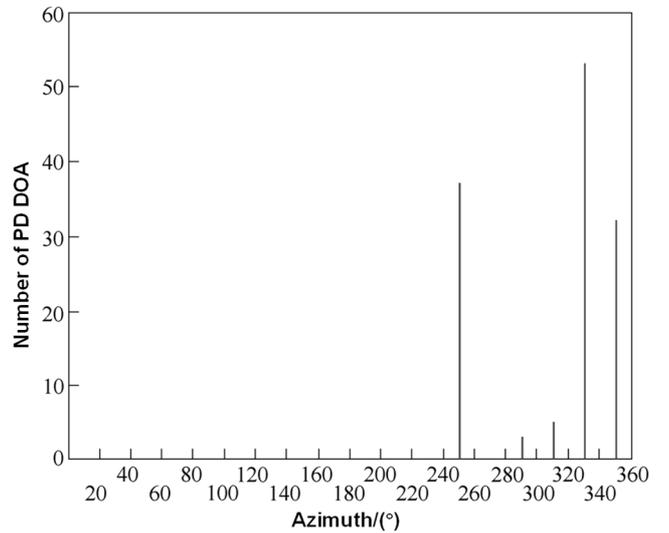

*Fig. 10.* Azimuth diagram of PD DOA results at 340° by MLE

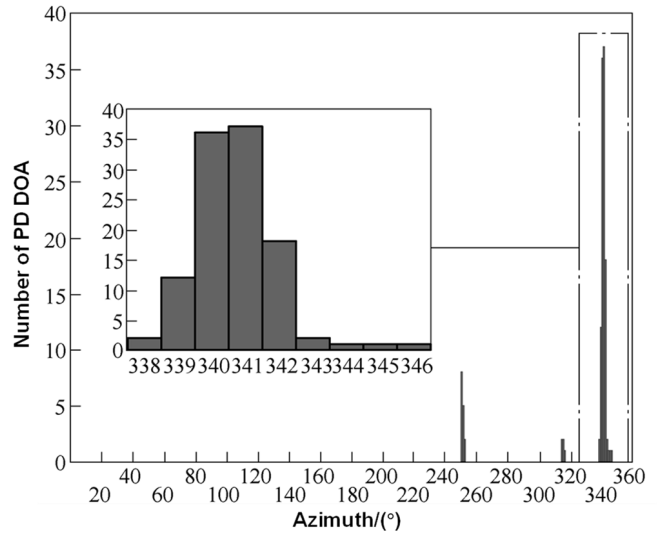

*Fig. 11.* Azimuth diagram of PD DOA results at 340° by MLE and spline interpolation

After the preliminary PD DOA results were obtained, the K-means algorithm was used to cluster the results. From Fig. 12 we can see that, the PD DOA results at 80° are classified into four clusters denoted by red boxes. The centre value of the largest cluster would be treated as the PD DOA results were closer to the actual azimuth angle. While other clusters are recognized as interference or noise.

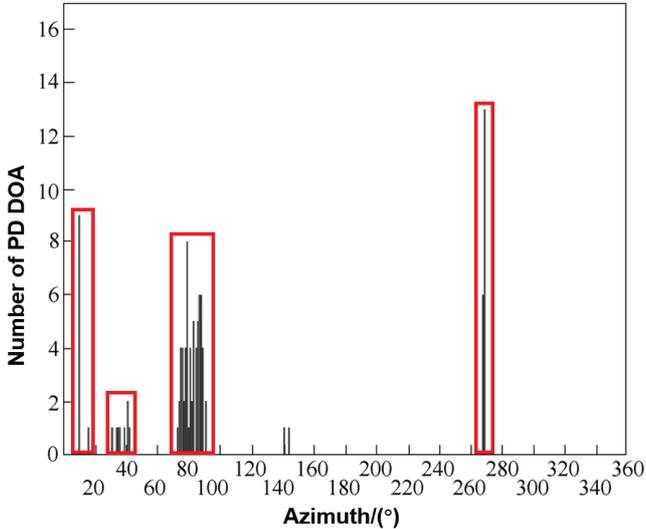

**Fig. 12.** *Illustration of the clusters of multiple PD DOA results at 80° after K-means algorithm*

The results of the four azimuth tests are summarized in Table 1. We can see that the average error of the DOA results obtained by only MLE is larger than that obtained by interpolation. In particular, for the 340° azimuth, the average error was approximately 12°. For the other three azimuth angles, the average errors were less than 6°. However, after the interpolation and clustering algorithm, the average errors for all four azimuth angles were less than 5°.

**Table 1.** PD DOA results of UHF four-sensor array

| No. | Azimuth / (°) | Method | Avg. of mean error / (°) |
|---|---|---|---|
| 1 | 40 | without interpolation | 2.17 |
|   |    | with interpolation | 2.87 |
| 2 | 80 | without interpolation | 5.87 |
|   |    | with interpolation | 4.08 |
| 3 | 150 | without interpolation | 2.38 |
|   |     | with interpolation | 2.03 |
| 4 | 340 | without interpolation | 11.72 |
|   |     | with interpolation | 0.71 |

Furthermore, the influence of the sensor number on the DOA accuracy was also investigated. The sensor number of the array increased from 4 to 12, and all the sensors were evenly placed on the tripod, as shown in Fig. 13. The azimuth angles for the PD DOA test were set as 10°, 150°, 270°, and 340°, respectively. Then, 120 tests were performed at each azimuth, as addressed before. Finally, the array was restored to four sensors, and the same tests were performed at the same azimuth angles for comparison.

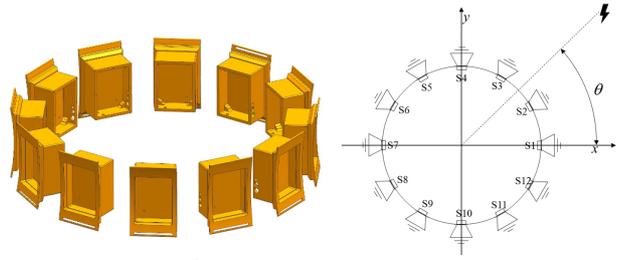

**Fig. 13.** *Diagram of 12-sensor array*

The results calculated by interpolation and clustering are reported in Table 2. Table 2 shows that, when the azimuth is 10°, the DOA error of the 12-sensor array was larger than that of the four-sensor array. For other azimuth angles, the difference in DOA error between these two types of array was not significant. Thus, in the tests that were conducted, the number of sensors had little effect on the DOA error. There was no statistical significance to show that the increase in the number of sensors was related to the improvement of DOA accuracy. However, more laboratory and field tests are needed to draw a more rigorous conclusion.

**Table 2.** DOA results of different number of UHF sensors

| Azimuth / (°) | DOA result (4 sensors) / (°) | Avg. of mean error (4 sensors) / (°) | DOA result (12 sensors) / (°) | Avg. of mean error (12 sensors) / (°) |
|---|---|---|---|---|
| 10 | 10.12 | 0.12 | 15.81 | 5.81 |
| 150 | 152.33 | 2.33 | 151.44 | 1.44 |
| 270 | 267.82 | 2.18 | 268.27 | 1.73 |
| 340 | 342.07 | 2.07 | 338.29 | 1.71 |

Considering that the distance between sensor array and PD source will affect RSSI measurement accuracies, the DOA results with different settings on distances (6 m & 12 m) are tested and reported in Table 3. It's shown that DOA errors may increase with increasing distances. Therefore, the maximum effective range of proposed method needs to be further investigated in practical applications.

**Table 3.** DOA results of different distances

| Azimuth / (°) | DOA result (6 m) / (°) | Avg. of mean error (6 m) / (°) | DOA result (12 m) / (°) | Avg. of mean error (12 m) / (°) |
|---|---|---|---|---|
| 10 | 10.12 | 0.12 | 11.57 | 1.57 |
| 150 | 152.33 | 2.33 | 153.71 | 3.71 |
| 270 | 267.82 | 2.18 | 275.14 | 5.14 |
| 340 | 342.07 | 2.07 | 343.22 | 3.22 |

### 4.2. Error Analysis

According to the theory of statistical signal processing, the variance of unbiased estimation has a theoretical lower limit, and its lower limit value can be used as one of the important indices to evaluate the performance of the location algorithm. The common lower bound of the Cramer–Rao lower bound (CRLB) [27] is

$$\mathrm{var}(\varGamma) \geqslant I^{-1}(\varGamma) \qquad (21)$$

where $I(\varGamma)$ is the Fisher information matrix. The variance estimation for $\theta$ in $\varGamma$ is

$$\mathrm{var}(\theta) \geqslant \left[I^{-1}(\varGamma)\right]_{1,1} \qquad (22)$$

$$\left[I(\varGamma)\right]_{1,1} = -E\left[\frac{\partial^2 \ln p(P_r;\varGamma)}{\partial \theta^2}\right] \qquad (23)$$

Substituting Eq. (16) into Eq. (23), one can obtain

$$\frac{\partial^2 \ln p(\boldsymbol{P}_r;\Gamma)}{\partial \theta^2} = \sum_{m=1}^{M} \frac{1}{2\sigma_m^4}\left(\frac{\partial \sigma_m^2}{\partial \theta}\right)^2 - \frac{1}{2\sigma_m^2}\cdot\frac{\partial^2 \sigma_m^2}{\partial \theta^2} -$$
$$\frac{(P_{r,m}-\mu_m)^2}{\sigma_m^6}\left(\frac{\partial \sigma_m^2}{\partial \theta}\right)^2 - \frac{1}{\sigma_m^2}\left(\frac{\partial \mu_m}{\partial \theta}\right)^2 - \quad (24)$$
$$\frac{2(P_{r,m}-\mu_m)}{\sigma_m^4}\cdot\frac{\partial \sigma_m^2}{\partial \theta}\cdot\frac{\partial \mu_m}{\partial \theta} +$$
$$\frac{(P_{r,m}-\mu_m)^2}{2\sigma_m^4}\cdot\frac{\partial^2 \sigma_m^2}{\partial \theta^2} + \frac{P_{r,m}-\mu_m}{\sigma_m^2}\cdot\frac{\partial^2 \mu_m}{\partial \theta^2}$$

Then,
$$[I(\Gamma)]_{1,1} = \sum_{m=1}^{M}\left[\frac{1}{2\sigma_m^4}\left(\frac{\partial \sigma_m^2}{\partial \theta}\right)^2 + \frac{1}{\sigma_m^2}\left(\frac{\partial \mu_m}{\partial \theta}\right)^2\right] \quad (25)$$

Thus, the CRLB is obtained by
$$\mathrm{CRLB}(\theta) = \left[I(\Gamma)^{-1}\right]_{1,1} \quad (25)$$

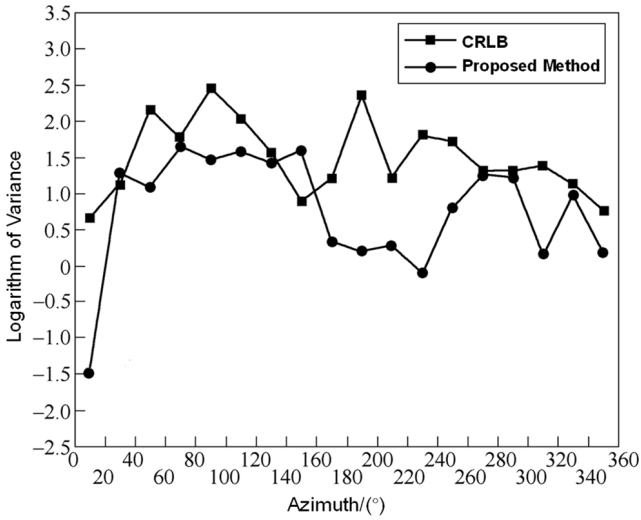

**Fig. 14.** *Variance of proposed method and theoretical CRLB*

Based on the theoretical analysis, the proposed method and the CRLB are compared in Fig. 14. It's noticed that, the DOA results of the proposed method are better than the lower bound of unbiased estimations other than for 30° and 150°. Thus, the proposed method has a better accuracy than the theoretical lower bound.

However, since only amplitudes and phases of PD are obtained by the designed UHF wireless sensors, many detailed characteristics (e.g. time domain characteristics, frequency information, etc.) might be lost, improvements on the analysing methods based on RSSI measurements are limited compared to traditional time-difference based methods. In addition, the accuracy would be further influenced by hardware design and ambient noise, considering difficulties in practical applications.

## 5. Comparison with Other Typical PD Localization Methods

A comprehensive comparison of accuracy, cost, and limitations between the proposed method and state-of-the-art PD localization techniques is shown in Table 4.

In particular, the localization accuracy reported in the most referenced research [8] states that, when PD sources are within 5 m of the UHF sensor array, the obtained localization accuracy is within 1 m. When the PD sources are at 12 m, the localization error is an excess of 2 m (or 5°). From a field application point of view, the distance between the sensor array and the power equipment in AIS frequently exceeds 30 m [8]. From the comparison listed in Table 4, the accuracy achieved by the proposed method is close to that of TDOA. However, because the UHF RSSI is adopted for PD localization, the sampling rate is reduced (GSa/s to MSa/s), which makes the proposed method more cost effective. Compared with the RSSI fingerprinting-based PD localization method, on the one hand, the fingerprinting map needs to be built by conducting a site survey. On the other hand, it is an online monitoring method so that, when more power equipment needs to be monitored, more sensors are needed [16]. Thus, this type of PD localization method is not sufficiently convenient for practical application. The proposed method and system could also be installed on the vehicle for mobility and ease of use.

**Table 4.** Comparison of different PD localization methods

| Method | Claimed accuracy | Cost | Extendibility | Complexity | Limitations |
|---|---|---|---|---|---|
| Time difference of signals (TDOS) | Up to decimetres | Medium, depends on the cost of oscilloscope | High, time domain waveform sampling, details of PD are preserved | Low | Professional operation by experts required |
| Time difference of arrival (TDOA) | Less than 2 m (or 5°) | High, depends on the high-speed synchronous acquisition device (GSa/s) | High, time domain waveform sampling, details of PD are preserved | High | Size and weight are large, usually installed on the vehicle |
| Received signal strength indicator (RSSI) fingerprinting | 90% less than 1.3 m | Low (MSa/s) | Low, only amplitude and phase of PD are obtained | Medium | Extra prebuilt fingerprinting map required |
| **Proposed method** | Less than 6° | Low (MSa/s) | Low, only amplitude and phase of PD are obtained | Medium | Performance needs to be verified in real substations |

## 6. Conclusion

The purpose of this study was to develop a PD DOA method and system for AISs that is low cost and easy to use. Therefore, a statistical approach using the PD DOA method was developed based on a UHF wireless sensor array and RSSI MLE algorithm. Theoretical derivation and laboratory tests were performed, and some conclusions can be drawn.

(1) The results of laboratory tests proved the feasibility of PD DOA using UHF RSSI data. However, the average DOA error is quite large when only the MLE is considered. The results indicate that, if interpolation and a clustering algorithm are applied, the maximum average error is less than 6°. Clustering, as a popular machine-learning algorithm, also plays an important role in PD localization, indicating that the data-driven analysis may help improve the PD localization accuracy.

(2) The laboratory test shows that the proposed method has a potential application in the identification of power equipment with insulation deterioration, but the accuracy and maximum effective range still needs to be verified further in actual air-insulated substations with complex ambient noise and disturbances.

(3) The focus was on the proposed statistical analysis framework for PD DOA in AISs; thus, only a single PD source was considered in this study. Examining the more important multiple-PD-source DOA is planned as future work. On the other hand, analyse on the distances between PD source and UHF sensor array via only RSSI measurements would be considered as a valuable supplement to the proposed DOA method in future work.

# 7. Acknowledgments

This work was supported in part by the National Key Research and Development Program (2017YFB0902705) and the State Grid Science and Technology Program of China.